\def\itp{ITP-SB-96-73}
\documentclass[12pt]{article}
\def\beq {\begin{equation}}
\def\eeq {\end{equation}}
\def\be {\begin{equation}}
\def\ee {\end{equation}}
\def\barr{\begin{array}}
\def\earr{\end{array}}
\def\bea{\begin{eqnarray}}
\def\eea{\end{eqnarray}}
\def\bmath{\begin{displaymath}}
\def\emath{\end{displaymath}}
\def\bq{\begin{quote}}
\def\eq{\end{quote}}
\def\oas{$O(\alpha_s)$}

\def\g5{\gamma_5}
\def\as{\alpha_s}
\def\real{\mathop{\mbox{\rm Re}}\nolimits}

\def\gE{\gamma_{\scriptscriptstyle E}}
\def\chiz{\chi_{\scriptscriptstyle Z}}
\def\mz{M_{\scriptscriptstyle Z}}
\def\gz{\Gamma_{\scriptscriptstyle Z}}
\def\gf{g_{\scriptscriptstyle F}}
\def\CF{C_{\scriptscriptstyle F}}

\def\ct{\cos\theta}
\def\c2t{\cos^2\kern-2pt\theta}
\def\st{\sin\theta}
\def\s2t{\sin^2\kern-2pt\theta}

\def\Li{\mbox{$\mbox{\rm Li}_2$}}
\def\Frac#1#2{\mbox{$\textstyle{#1\over#2}$}}
\def\half{\Frac{1}{2}}
\def\eps{\varepsilon}

\def\MZ{M_{\scriptscriptstyle Z}}

\def\Slash#1{\mbox{$\not{\hspace{-1.03mm}#1}$}}

\def\direc{\bf{\hat{p}}}
\def\ndirec{\bf{\hat{n}}}
\def\I#1{{I}_{#1}}
\def\S#1{{S}_{#1}}

\def\K#1{{K}_{#1}}

\def\KK#1{{\tilde{K}}_{#1}}

\def\kint{\int{dy\,dz\over\sqrt{(1-y)^2-\xi}\sqrt{(1-z)^2-\xi}\:\:}\kern.1cm}
\def\sxi{\mbox{$\sqrt\xi$}}

\begin{document}
\vspace*{-2.8cm}
\begin{flushright}
hep-ph/9706444 \\ [-.2cm]
\itp \\ [-.2cm]
FTUV-97/39 \\ [-.2cm]
IFIC-97/39 \\ [-.2cm]
\end{flushright}
\begin{center}
{\Large\bf{\boldmath $O(\alpha_s)$} Spin-Spin Correlations for Top and \\[.25cm]
         Bottom Quark Production in {\boldmath $e^+e^-$} Annihilation}
\vskip.75cm
M.~M.~Tung$^a$, J.~Bernab\'eu$^b$, J.~Pe\~narrocha$^b$
\vskip.4cm
\it{$^a$Institute for Theoretical Physics\\
State University of New York\\
Stony Brook, NY 11794-3840, U.S.A}
\vskip.25cm
\it{$^b$Instituto de F\'\i sica Corpuscular, Departament de F\'\i sica Te\`orica\\
Universitat de Val\`encia, 46100 Burjassot (Val\`encia), Spain}\\
\vskip2.5cm
{\bf Abstract}
\end{center}
\vskip.1cm
We present the full \oas\ longitudinal spin-spin correlations for heavy-quark
pair production at $e^+ e^-$ high-energy colliders in closed analytical form.
In such reactions, quark and antiquark have strongly correlated spins, and
the longitudinal components are dominant. For the explicit computation of the
QCD bremsstrahlung contributions, new phase-space integrals are derived.
Explicit numerical estimates are given for $t\bar{t}$ and $b\bar{b}$
production. Around the $Z$-peak, QCD one-loop corrections depolarize the
spin-spin asymmetry for bottom quark pairs by approximately $-$4\%.
For top pair production, we find at 350~GeV a 0.6\% increased polarization
over a value of 0.4 in the longitudinal correlation. For more than 1~TeV the
\oas\ corrections enhance depolarization to $-$2\% in the top-pair case.
\newpage
After the long-sought discovery of the top quark at the Fermilab
Tevatron~\cite{CDF}, now much effort is made to further investigate
all the properties of this exceptional particle, which completes the
missing SU(3) ingredient in the third family of the Standard Model.
A thorough experimental analysis of the top quark in combination with
theoretical predictions offers unique possibilities to explore so-far
unaccessible domains of the Standard Model.

Much work has gone into the study of polarized top production from hadron
colliders~\cite{brand} as these machines more easily provide the necessary
energies to produce the heavy particle. However, once $e^+ e^-$ colliders have
reached the required production threshold and beyond, many new interesting
experimental tests may be carried out and add to the wealth of existing data
on the top~\cite{jezabek,schmidt}. In particular, polarized top production
from $e^+ e^-$ annihilation deserves special attention. Not only are the spin
properties of the top in the final state fully accessible through its rapid
and predominant electroweak decay mode, but additionally $e^+ e^-$ collisions
provide a clean initial state from the experimental and theoretical standpoint.

Heavy-quark production from high-energy $e^+ e^-$ collisions, especially
top production, will permit high-precision measurements to directly probe
the respective photon and $Z$-boson interaction vertices.
In these processes, the longitudinal polarization $P_L$ of quark or
antiquark is large, whereas the remaining transverse components contribute
to a smaller extent~\cite{trans,ps}. QCD one-loop corrections for the
longitudinal polarization averaged over the scattering angle have been 
calculated before and change $\langle P_L\rangle$ typically by 3\% for the
top~\cite{kpt}. Other \oas\ single-spin asymmetries have also been 
discussed~\cite{gkt}, using the techniques developed in Ref.~\cite{kpt}.
A recent work by Olsen and Stav~\cite{os} investigates quark-pair
polarization effects in 2- and 3-jet events from polarized $e^+ e^-$
beams.

In the present paper, we shall concentrate on the massive QCD one-loop
corrections for the total spin-spin asymmetry $\langle P_{LL}\rangle$, which
measures the bilinear spin correlations among the longitudinal quark and
antiquark polarizations. The relevant observable is defined as follows
\be\label{PLL}
\langle P_{LL}\rangle =
{\sigma_{tot}(\uparrow\uparrow)
-\sigma_{tot}(\uparrow\downarrow)
-\sigma_{tot}(\downarrow\uparrow)
+\sigma_{tot}(\downarrow\downarrow)\over4\,\sigma_{tot}},
\ee
where $\sigma_{tot}$ denotes the total cross section for
$e^+e^-\to\gamma,Z\to q\bar{q}$, and $\uparrow$, $\downarrow$ refer to
positive and negative helicities for quark and antiquark, respectively.
It is straightforward to see that in the numerator all combinations
of $({\bf I}+{\bf s_1})\otimes({\bf I}+{\bf s_2})$ except for the
bilinear spin form ${\bf s_1}\otimes{\bf s_2}$ drop out. Note that the
normalization factor of $1/4$ is an arbitrary but convenient choice in
order to obtain for massless quarks at Born level $P_{LL}^{m=0}(Born\/)=1$,
due to chiral current conservation.
 
For the calculation of the spin-dependent cross sections, we group the
hadronic tensor $H^{\mu\nu}$ into separate parity-parity combinations
$i,j=V,A$, so that the structure of vector and axial-vector interference
terms becomes more transparent. Then, for a fixed combination $ij$ each 
contraction with the lepton tensor $L^{\mu\nu}$ has to be multiplied by 
the corresponding compound coupling $g^{ij}$, which contains products
of the usual neutral-current couplings $v_f=2T^f_z-4Q_f\sin^2\theta_W$ and 
$a_f=2T^f_z$, and the fractional fermion charge $Q_f$. Finally, integration
over the azimuthal $(\varphi)$ and polar $(\theta)$ angles of the quark
(including the flux factor and $\gamma,Z$ propagator term) yields
\be\label{master}
\sigma_{tot} = {N_C\over16} \left({\alpha\over q^2}\right)^2 v
               \sum_{i,j=V,A} g^{ij}\int d\varphi \int d \cos\theta\:
               L^{\mu\nu} H^{ij}_{\mu\nu},
\ee
where $N_C=3$ accounts for the color-singlet state of the produced quark pair, 
and $v=\sqrt{1-4m^2/q^2}$ is the common Schwinger mass parameter ($m$ quark
mass, $q$ total energy-momentum transfer).

In order to directly determine the total cross section, it is practical to 
introduce the following covariant projection operator~\cite{kpt}
$$
\Pi^{\mu\nu}_T\sim\left(g^{\mu\nu}-{q^\mu q^\nu\over q^2}\right),
$$
so that at QCD one-loop level Eq.~(\ref{master}) simplifies to
\be\label{master2}
\sigma_{tot} = {3\over8}\pi\left({\alpha\over q^2}\right)^2 v
               \sum_{i,j=V,A} g^{ij} \left[H^{(2)ij}_T+H^{(3)ij}_T\right].
\ee
Here, it is $H_T\equiv\Pi^{\mu\nu}_T H_{\mu\nu}$, and the additional superscript
separates the virtual (including the Born term) from the real contributions.
Note that a relative factor between two- and three-particle phase-space has to
be taken into account~\cite{tbp}.

In Eq.~(\ref{master2}), the explicit expressions for $g^{i,j}$ are
\bea
g^{VV}   &=&
Q_q^2-2\,Q_q v_e v_q \real{\chiz}+(v_e^2+a_e^2)\,v_q^2\,|\chiz|^2, \\
g^{AA}   &=& (v_e^2+a_e^2)\,a_q^2\,|\chiz|^2, \\
g^{V\!A} &=& -Q_q a_e a_q \real{\chiz}+2\,v_e a_e v_q a_q |\chiz|^2,
\eea
and we consider the finite width of the $Z$ boson by taking
\be
\chiz(q^2)={\gf\,\mz^2\,q^2\over q^2-\mz^2+i\mz\,\gz}\quad\mbox{with}\quad
\gf={G_{\scriptscriptstyle F}\over 8\sqrt{2}\,\pi\alpha}\approx
4.299\cdot 10^{-5}\,\mbox{GeV}^{-2}.
\ee

For the explicit spin-dependent computation of $H^{(3)}_T$ in 
Eq.~(\ref{master2}), we use the following kinematical framework. As usual
the direction of the outgoing quark in the center-of-momentum system (cms)
is given by
\be
\direc = \bf{p_1}/|\bf{p_1}|
       = \Big(\st\,\cos\varphi,\,\st\,\sin\varphi,\,\ct\,\Big).
\ee
Then, the vector normal to the production plane is
\be
\ndirec = \Big(-\ct\,\cos\varphi,\,-\ct\,\sin\varphi,\,\st\,\Big),
\ee
and all outgoing momenta in the reaction 
$e^+e^-\to q(p_1)\,\bar{q}(p_2)\,g(p_3)$
are confined to within this plane. Moreover, given also the direction of 
the antiquark $(\bf{\hat{p}_2})$ the gluon direction $(\bf{\hat{p}_3})$ is
fully determined by energy-momentum conservation. Thus, for the parametrization
of the three-particle phase space the following two dimensionless energy scales
suffice
\be
y=1-{2\,p_1\cdot q\over q^2}, \hskip1cm z=1-{2\,p_2\cdot q\over q^2}.
\ee
In fact, it is this choice of phase-space parametrization in combination with 
sophisticated integration techniques that allows for a complete analytical
evaluation of $H^{(3)}_T$. The method was first devised in Refs.~\cite{kpt}.

The longitudinal spin components of quark and antiquark are now given
respectively by\footnote{The choice of metric is standard:
$g^{\mu\nu}=\mbox{\rm diag\/}(1;\,-1,-1,-1)$.}
\bea
s_1 &=& 2\lambda_1
\xi^{-{1\over2}}\Big(\sqrt{(1-y)^2-\xi};\ \direc\,(1-y)\Big), \\[.25cm]
s_2 &=& 2\lambda_2
\xi^{-{1\over2}}\Big(\sqrt{(1-z)^2-\xi};\ \bf{\hat{p}_2}(1-z)\Big),
\eea
where $\lambda_1, \lambda_2=\pm\half$ are the corresponding helicities, and 
$\xi=1-v^2=4m^2/q^2$. It is now straightforward to make the appropriate
substitutions in the gamma traces $H^{(3)}_T$, and use
\bea
u(p_1,s_1)\,\bar{u}(p_1,s)&\longrightarrow&
\half\Big(1+\g5\Slash{s_1}\Big)\Big(\Slash{p_1}+m\Big), \\
v(p_1,s_1)\,\bar{v}(p_1,s)&\longrightarrow&
\half\Big(1+\g5\Slash{s_2}\Big)\Big(\Slash{p_2}-m\Big).
\eea
Here, there are no technical difficulties in the proper treatment of the
axial-vector current within any renormalization scheme, since the product 
${\bf s_1}\otimes{\bf s_2}$ implies an even number of $\g5$-matrices.
Dimensional reduction, {\it i.e.} the Clifford algebra of the Dirac
matrices is performed in four dimensions, is the simplest renormalization
procedure to control the UV sector~\cite{dimred,kt}, and we shall adopt
it in the subsequent calculations.

Note that in the previous discussion, two-particle kinematics is fully
recovered by putting $y=z=0$, that is, both quark and antiquark share
an equal amount of the total available cms Energy $E_{cms}^2=q^2=(p_1+p_2)^2$.

The individual \oas\ virtual corrections to the Born approximation in
Eq.~(\ref{master2}) are now derived in terms of the charge and magnetic
moment form factors $A$, $B$, and $C=A-2B$ (see {\it e.g.} Ref.~\cite{tbp}),
where the last identity is an immediate consequence of axial-vector current
conservation in the massless limit. Our explicit results are
\bea
& H^{(2)VV}_T &=\ \Frac{16}{3}q^2\Big[\,\Big\{2(1-4\lambda_1\lambda_2)
                  +\xi(1+4\lambda_1\lambda_2)\Big\}(1+2A)\label{virt1}
              \nonumber \\
&             & \hskip2cm +2(1+4\lambda_1\lambda_2)v^2B\,\Big], \\
& H^{(2)AA}_T &=\ \Frac{32}{3}q^2 v^2 (1-4\lambda_1\lambda_2)(1+2C), \\
& \half\Big(H^{(2)V\!A}_T+H^{(2)AV}_T\Big)\label{virt3}
              &=\ \Frac{64}{3}q^2 v(\lambda_1-\lambda_2)(1+A+C).
\eea
It is important to notice that the form factors contain soft divergences,
which have their origin in the massless nature of the gluon fields. In these
expressions, the dimensional regulator $\eps=4-N$ is equivalent to an
infinitesimal gluon mass cut-off $\Lambda=m_g^2/q^2$. This equivalence is
readily established by the correspondence rule~\cite{muta,tbp}
\be\label{corrule}
\ln\Lambda\,\longleftrightarrow\,{2\over\eps}-
\gE+\ln\left({4\pi\mu^2\over q^2}\right).
\ee
These at most logarithmic divergences in $\Lambda$ also emerge after
integrating out the following differential distributions of the \oas\
tree-graph contributions in Eq.~(\ref{master2}):
\bea\label{bl1}
{dH^{(3)VV}_T\over dy\ dz\ \ } &=& 
{\as\over\pi}\,\CF{8\over3}{q^2\over v}\Bigg[
-2(2+\xi)\left({1\over y}+{1\over z}\right)-\half\xi(2+\xi)\left(
{1\over y^2}+{1\over z^2}\right)+2\,{y\over z}+2\,{z\over y}
\nonumber \\ &&
+(4-\xi^2){1\over yz}
+{2\lambda_1\lambda_2\over\sqrt{(1-y)^2-\xi}\,\sqrt{(1-z)^2-\xi}}\,
\Bigg\{-4(2-\xi)(3-\xi)
\nonumber \\ &&
+(2-\xi)(8-7\xi)\left({1\over y}+{1\over z}\right) 
+\xi(1-\xi)(2-\xi)\left({1\over y^2}+{1\over z^2}\right)
\nonumber \\ &&
-2(1-\xi)(2-\xi)^2{1\over yz}-4(y^2+z^2)
+4(1-\xi)\left({y^2\over z}+{z^2\over y}\right)
\nonumber \\ &&
+\xi(2-\xi)\left({y^2\over z^2}+{z^2\over y^2}\right)
+4(3-2\xi)(y+z)+(-12+10\xi-3\xi^2)\left({y\over z}+{z\over y}\right)
\nonumber \\ &&
-\xi(2-\xi)\left({y\over z^2}+{z\over y^2}\right)
\,\Bigg\}\ \Bigg], \\[.75cm]
{dH^{(3)AA}_T\over dy\ dz\ \ } &=& \label{bl2}
{\as\over\pi}\,\CF{8\over3}{q^2\over v}\Bigg[\,
2\xi-4(1-\xi)\left({1\over y}+{1\over z}\right)
-\xi(1-\xi)\left({1\over y^2}+{1\over z^2}\right)
\nonumber \\ &&
+(2+\xi)\left({y\over z}+{z\over y}\right)
+2(1-\xi)(2-\xi){1\over yz}
\nonumber \\ &&
+{2\lambda_1\lambda_2\over\sqrt{(1-y)^2-\xi}\,\sqrt{(1-z)^2-\xi}}\,
\Bigg\{-4(6-5\xi)
+2(1-\xi)(8-7\xi)\left({1\over y}+{1\over z}\right)
\nonumber \\ &&
+2\xi(1-\xi)^2\left({1\over y^2}+{1\over z^2}\right)
-4(1-\xi)^2(2-\xi){1\over yz}-2(2-\xi)(y^2+z^2)
\nonumber \\ &&
+2(2-3\xi)\left({y^2\over z}+{z^2\over y}\right)
+2\xi(1-\xi)\left({y^2\over z^2}+{z^2\over y^2}\right)
+2(6-\xi)(y+z)
\nonumber \\ &&
-2(6-8\xi+\xi^2)\left({y\over z}+{z\over y}\right)
-2\xi(1-\xi)\left({y\over z^2}+{z\over y^2}\right)
+4\xi\,yz\,\Bigg\}\ \Bigg],
\eea
\bea\label{linear}
{d\Big(H^{(3)V\!A}_T+H^{(3)AV}_T\Big)\over2\ dy\ dz\ \ } &=&
{\as\over\pi}\,\CF{16\over3}{q^2\over v}
{\lambda_1\over\sqrt{(1-y)^2-\xi}}\,
\Bigg[\,
4-\xi
-(4-5\xi){1\over y}-2(4-3\xi){1\over z} 
\nonumber \\ &&
-\xi(1-\xi)\left({1\over y^2}+{1\over z^2}\right)
+\xi\left({y\over z^2}-{y^2\over z^2}\right)
-2z+(2-\xi){z\over y}
\\ &&
+(6-\xi){y\over z}-2{y^2\over z}+2(1-\xi)(2-\xi){1\over yz}
\ \Bigg]-\Big(\lambda_1\to\lambda_2,y\leftrightarrow z\Big).
\nonumber
\eea
In the last expression $\Big(\lambda_1\to\lambda_2,y\leftrightarrow z\Big)$
denotes an additional term linear in the antiquark helicity $\lambda_2$, and
$y$ and $z$ are interchanged. For completeness, we have presented the full
differential distributions displaying all possible spin contributions.
The first term of Eq.~(\ref{linear}) agrees with the result previously derived
for longitudinal quark polarization~\cite{kpt}. Including the second term,
Eq.~(\ref{linear}) becomes $C$-odd under the exchange of quark and antiquark
in the final state, so that bilinear spin terms can not occur. On the
other hand, Eqs.~(\ref{bl1}) and (\ref{bl2}) are $C$-even, and thus
contain only unpolarized and bilinear spin terms. Note that angular
extensions of these distributions can be found in Ref.~\cite{os} in
different form.

We perform the integration of Eqs.~(\ref{bl1})--(\ref{linear}) over the
three-body phase in fully analytical form. Apart from the unpolarized and
linear spin terms (which comprise known integrals of the $I$- and $S$-type,
respectively~\cite{kpt}), entirely new phase-space integrals have to be
solved for the bilinear spin contributions. In this case, phase-space
symmetry is restored, and this essentially reduces the amount of integrals
to be tackled by half. However, the analytical evaluation is considerably
more complicated due to the additional spin-spin weight
$\Big[\sqrt{(1-y)^2-\xi}\sqrt{(1-z)^2-\xi}\Big]^{-1}$.
These new integrals are classified in the Appendix along with a listing of
their explicit solutions. We shall henceforth refer to them as type 
$K$-integrals. Upon integration, one then obtains

\bea
H^{(3)VV}_T &=&
{\as\over\pi}\,\CF{8\over3}{q^2\over v}\Bigg[\,
-4(2+\xi)\I{2}-\xi(2+\xi)\I{3}+4\I{4}+(4-\xi^2)\I{5}
\nonumber \\ &&
+4\lambda_1\lambda_2\Big\{-2(2-\xi)(3-\xi)\K{1}+(2-\xi)(8-7\xi)\K{2} 
\nonumber \\ &&
+\xi(1-\xi)(2-\xi)\K{3}-(1-\xi)(2-\xi)^2\K{4}-4\K{5}+4(1-\xi)\K{6}
\nonumber \\ &&
+\xi(2-\xi)\K{7}+4(3-2\xi)\K{8}+(-12+10\xi-3\xi^2)\K{9}
\nonumber \\ && \label{real1}
-\xi(2-\xi)\K{10}\,\Big\}\ \Bigg], \\[.75cm]
H^{(3)AA}_T &=&
{\as\over\pi}\,\CF{8\over3}{q^2\over v}\Bigg[\,
2\xi\I{1}-8(1-\xi)\I{2}-2\xi(1-\xi)\I{3}+2(2+\xi)\I{4}
\nonumber \\ &&
+2(1-\xi)(2-\xi)\I{5}+4\lambda_1\lambda_2\Big\{-2(6-5\xi)\K{1}
+2(1-\xi)(8-7\xi)\K{2}
\nonumber \\ &&
+2\xi(1-\xi)^2\K{3}-2(1-\xi)^2(2-\xi)\K{4}-2(2-\xi)\K{5}+2(2-3\xi)\K{6}
\nonumber \\ &&
+2\xi(1-\xi)\K{7}+2(6-\xi)\K{8}-2(6-8\xi+\xi^2)\K{9}-2\xi(1-\xi)\K{10}
\nonumber \\ && 
+2\xi\K{11}\,\Big\}\ \Bigg], \label{real2}
\eea
\bea
\half\Big(H^{(3)V\!A}_T+H^{(3)AV}_T\Big) &=&
{\as\over\pi}\,\CF{16\over3}{q^2\over v}(\lambda_1-\lambda_2)\Bigg[\,
(4-\xi)\S{1}-(4-5\xi)\S{2}-2(4-3\xi)\S{4} 
\nonumber \\ &&
-\xi(1-\xi)(\S{3}+\S{5})+\xi(\S{6}-\S{7})-2\S{8}+(2-\xi)\S{9}
\nonumber \\ &&
+(6-\xi)\S{10}-2\S{11}+2(1-\xi)(2-\xi)\S{12}\,\Big\}\ \Bigg].
\label{real3}
\eea

We are now in the position to obtain the complete \oas\ results for
each individual parity-parity combination by adding the corresponding
virtual and real parts, Eqs.~(\ref{virt1})--(\ref{virt3}) and
Eqs.~(\ref{real1})--(\ref{real3}), respectively. All logarithmic
soft divergences cancel in the total sum, as we expect. Finally,
Eq.~(\ref{master2}) gives the correct normalization and flux factors,
and in the total spin correlation $\langle P_{LL}\rangle$, {\it viz.}
Eq.~(\ref{PLL}), only $\lambda_1\lambda_2$ terms originating from
$H^{VV}_T$ and $H^{AA}_T$ survive.

In Figure 1(a), we have plotted numerical estimates of
$\langle P_{LL}\rangle$ for top pair production depending on the
cms energy $E_{cms}=\sqrt{q^2}$. In all plots the QCD one-loop
corrections are indicated by dashed lines, the Born approximations
by solid lines. Figure 1(b) gives the denominator term of 
$\langle P_{LL}\rangle$, that is the total unpolarized cross section.
Figure 1(c) displays the impact that the \oas\ corrections have on
the Born approximation of $\langle P_{LL}\rangle$. We give
$\langle P_{LL}(O(\alpha_s))\rangle/\langle P_{LL}(Born\/)\rangle-1$
in percent, where $P_{LL}(O(\alpha_s))$ comprises the QCD one-loop
corrections as well as the Born contribution. Figures 2 present the
results for $e^+ e^-\to\gamma,Z\to b\,\bar{b}(g)$ in a similar
fashion. The initial boundary condition for the running of the strong
coupling constant is chosen to be $\alpha_s^{(5)}(\MZ)=0.123$ for five
active quark flavors.

Close to threshold, the QCD one-loop corrections significantly enhance
the unpolarized and polarized cross sections of the top quark pairs by 40\%
to 45\%. We wish to emphasize that in the region where $t$ and $\bar{t}$ are
at sufficiently small distances together, a QCD Coulomb interaction
is required to describe a correct threshold behavior~\cite{fks}.
For polarized top production the Green function method can be
used~\cite{jezabek,jezabek2}.

In Figure 1(a), the \oas\ $t\bar{t}$ spin correlations range between
$\langle P_{LL}\rangle=0.4$ at 350~GeV and $\langle P_{LL}\rangle=0.92$
at 1~TeV, which include \oas\ mass effects of 0.6\% and $-$0.7\%,
respectively. Above the 1~TeV limit, the total impact of the \oas\ corrections
to $\langle P_{LL}\rangle$ amounts to $-$1.5\% to $-$2\%.

In the bottom case, the QCD corrections produce a significant depolarization
around the $Z$-peak (Fig.~2(a)), opposite to the effect in the total rate
(Fig.~2(b)). In the 80--100~GeV range, the \oas\ corrections for
$\langle P_{LL}\rangle$ vary only slightly around $-$4\% (Fig.~2(c)). 

For very high cms energies, {\it i.e.} $\xi\to0$, the value
for $\langle P_{LL}\rangle$ at \oas\ stabilizes around 0.96, which differs from
the expected Born value of 1 by roughly $-$4\%.

The same effect is observable for up-type quarks, where this correction
equally amounts to approximately $-$4\%. The origin of this effect are
collinear helicity-flip terms of the order $m^2/\xi$. They do not emerge in the
naive computation, where the quark mass is set to zero in the very beginning.
Specifically, integrals $\K{3}$ and $\K{7}$ contain these IR/M singularities
and get multiplied by $\xi$ in Eqs.~(\ref{real1}) and (\ref{real2}). Similar 
finite corrections emerge in the massless limit for the longitudinal
single-spin polarization~\cite{kpt}. For a general discussion of this
phenomenon we refer to Ref.~\cite{fs}.

In summary, we have calculated the \oas\ radiative mass corrections to the 
total longitudinal spin correlations of heavy quark pairs produced in
$e^+ e^-$ annihilation. Fully analytical expressions have been obtained and
used for numerical estimates for top and bottom production. The impact
of these QCD one-loop corrections on the lowest Born approximation is
considerable, and we believe that future precision measurements of such
spin correlations at high-energy $e^+ e^-$ colliders help to reveal 
interesting new aspects of the Standard Model, or beyond. 
\vskip1cm
{\bf Acknowledgements.}  This work has been supported by the CICYT under
Grant No.\ AEN-96/1718 and the IVEI under Grant No.\ 96/036. M.M.T.\
gratefully acknowledges support by the Max-Kade Foundation, New York, NY,
and wishes to thank J.G.~K\"orner, J.~Smith, G.~Sterman, and S.~Groote for
useful comments.
\newpage
\leftline{\bf\Large Appendix: Analytical spin-spin phase-space integrals}
\vskip1cm
In terms of the kinematical variables $y=1-2p_1\cdot q/q^2$ and 
$z=1-2p_2\cdot q/q^2$ the three-body phase-space is confined to a region
within the limits~\cite{kpt}
\bea
y_- &=& \Lambda^{1\over2}\sxi+\Lambda, \quad y_+ = 1-\sxi, \\
z_\pm &=& {2y\over4y+\xi}\left[\,1-y-\Frac{1}{2}\xi+\Lambda+{\Lambda\over y}
          \pm {1\over y}\sqrt{(1-y)^2-\xi}\sqrt{(y-\Lambda)^2-\Lambda\,\xi}\, 
          \right],
\eea
where $\Lambda=m_g^2/q^2$ relates to a spurious gluon mass that regulates the
soft IR divergences in the intermediate results for the individual real and
virtual contributions, {\it viz.} Eq.~(\ref{corrule}).

Apart from the integral classes presented in previous \oas\
calculations (for a full listing see Refs.~\cite{tbp} and \cite{rt})
, new integral types emerge here in the bilinear
spin terms. The additional spin-spin weight 
$\Big[\sqrt{(1-y)^2-\xi}\sqrt{(1-z)^2-\xi}\Big]^{-1}$
complicates the computation of the explicit analytic integral solutions.
However, full phase-space symmetry $(y\leftrightarrow z)$ reduces the number
of integrals to be tackled.

In the calculation of the longitudinal spin-spin correlations for $e^+e^-\to
q(\uparrow)\,\overline{q}(\uparrow)\,g$ the following new class of analytic
integral expressions had to be derived:

     \begin{eqnarray}
     \K{1} & = & \kint \nonumber \\[.25cm]
     & = &
     \Li\left({1-v\over2}\right)-\Li\left({1+v\over2}\right)-
     4\Li\left(\Frac{1}{2}\sqrt{\xi}\right)
     \nonumber \\ &&
     +\left(-\Frac{1}{2}\ln\xi+\ln2
     \right)\ln\left({1+v\over1-v}\right)
     \nonumber \\ &&
     -\Frac{1}{2}\Big(\ln\xi-4\ln2\Big)
     \ln\xi+\Frac{1}{3}\pi^2-2\ln^2 2
     \nonumber \\
     \nonumber \\
     v\,\K{2} & = & v\,\kint {1\over y} \nonumber \\
     & = &
     2\Li\left({2-\sxi\over1+v}\right)+2\Li\left({1-v\over2-\sxi}\right)-
     4\Li\left(\sqrt{1-v\over1+v}\:\right)
     \nonumber \\ &&
     +\ln\left({1+v\over1-v}\right)
     \left[\,\ln\xi-2\ln(2-\sxi)+\ln\left({1+v\over1-v}\right)\,\right]
     \nonumber \\[.25cm] &&
     +\ln\xi\left[\,\Frac{1}{4}\ln\xi-\ln(2-\sxi)\,\right]+\ln^2(2-\sxi)
     \nonumber \\
     \nonumber \\
     \K{3} & = & \kint {1\over y^2} \nonumber \\
     & = &
     {4\over v\,\xi}\left[\,-\ln\Lambda^{1\over2}+\Frac{1}{2}\ln\xi
     -2\ln(2-\sxi)+2\ln v+2\ln2\,\right]
     \nonumber \\ &&
     +{2\over v^3}\left[\,\Li\left({2-\sxi\over1+v}\right)
     +\Li\left({1-v\over2-\sxi}\right)-\Li\left(\sqrt{1-v\over1+v}\:
     \right)-\Li\left({\sxi\over1+v}\right)\,\right]
     \nonumber \\ &&
     +{1\over v^3}\ln\left({1+v\over1-v}\right)\left[\,
     \ln\xi-2\ln(2-\sxi)-{2v\over\xi}\left\{(1+v)^2+1\right\}+
     \ln\left({1+v\over1-v}\right)\,\right]
     \nonumber \\ &&
     +{1\over v^3}\ln\xi\left[\,
     \Frac{1}{4}\ln\xi-{v(4-\xi)+6(1-\xi)\over\xi}\,\right]
     \nonumber \\ &&
     +{1\over v^3}\ln(2-\sxi)\left[\,-\ln\xi+\ln(2-\sxi)+2{v(4-\xi)+
     4(1-\xi)\over\xi}\,\right]
     \nonumber \\
     \nonumber \\
     \K{4} & = & \kint {1\over yz} \nonumber \\
     & = &
     {1\over v}\S{12}-{1\over v^2}\Bigg\{\;2\left[\,
     -\Li\left({1-v\over1+v}\right)+\Li\left({2-\sxi\over1+v}\right)
     \right. \nonumber \\ && \left.
     +\Li\left({1-v\over2-\sxi}\right)-\Li\left({1+v-\sxi\over2\,v}\right)
     +\Li\left(-{1-v-\sxi\over2\,v}\right)\,\right]
     \nonumber \\ &&
     +\ln\left({1+v\over1-v}\right)\left[\ln\xi-2\ln(2-\sxi)-\ln v
     +\Frac{3}{4}\ln\left({1+v\over1-v}\right)-\ln2\,\right]
     \nonumber \\ &&
     +\ln^2\Big(1+v-\sxi\Big)-\ln^2\Big(-(1-v-\sxi\,)\Big)
     +\ln^2(2-\sxi)
     \nonumber \\ &&
     +\ln\xi\left[\,\Frac{1}{4}\ln\xi-\ln(2-\sxi)\,\right]
     -\Frac{1}{3}\pi^2\;\Bigg\}
     \nonumber \\
     \nonumber \\
     \K{5} & = & \kint y^2 \nonumber \\
     & = &
     2\K{8}-\K{1}
     \nonumber \\ &&
     +\Frac{1}{2}\xi\left[\,
     \Li\left({1-v\over2}\right)-\Li\left({1+v\over2}\right)+
     2\Li\left(1-\Frac{1}{2}\sqrt{\xi}\right)-
     2\Li\left(\Frac{1}{2}\sqrt{\xi}\right)\,\right]
     \nonumber \\ &&
     +\ln(2-\sqrt{\xi})\left[\,\Frac{1}{8}\xi^2-\Frac{1}{2}\xi(2\ln2-\ln\xi)
     -1\,\right]
     \nonumber \\ &&
     +\Frac{1}{4}\ln\xi\left[\,\Frac{1}{4}\left(1-\xi^2\right)
     +\xi(2\ln2-\ln\xi)+\Frac{7}{4}\,\right]
     \nonumber \\ &&
     +\Frac{1}{2}\ln\left({1+v\over1-v}\right)\left[\,
     -\Frac{1}{8}\xi^2+\Frac{1}{2}\xi(2\ln2-\ln\xi)+1\,\right]
     \nonumber \\ &&
     +\Frac{1}{8}\xi^{3\over2}
     -{\xi^2\over8(2-\sqrt{\xi})}+\Frac{3}{2}(1-\sqrt{\xi})
     -\Frac{1}{8}v(7-v^2)
     \nonumber \\
     \nonumber \\
     \K{6} & = & \kint {y^2\over z} \nonumber \\
     & = &
     \left(1+\Frac{1}{2}\xi\right)\K{2}-3\ln\left({1+v\over1-v}\right)
     -\Frac{3}{2}\ln\xi+3\ln(2-\sqrt{\xi})
     \nonumber \\ &&
     +{\xi\over2(2-\sqrt{\xi})}-\Frac{3}{2}\sqrt{\xi}+1
     \nonumber \\
     \nonumber \\
     \K{7} & = & \kint {y^2\over z^2} \nonumber \\
     & = &
     \left(1+\Frac{1}{2}\xi\right)\KK{3}
     +{2v\over\xi}\left(1-{3\over v^2}\right)\left[\,
     -\Frac{1}{2}\ln\xi+\ln2-1\,\right]
     \nonumber \\ &&
     +\left({8+\xi\over2\,\xi}+{2+\xi\over v\,\xi}\right)\ln\xi
     +\left({6+\xi\over\xi}+2{2+\xi\over v\,\xi}\right)
      \ln\left({1+v\over1-v}\right)
     \nonumber \\ &&
     -{8+\xi\over\xi}\ln(2-\sqrt{\xi})
     -2{2+\xi\over v\,\xi}\left[2\ln v+\ln2\right]
     -{4\over\sqrt{\xi}}-{2v\over\xi}-{2\over2-\sqrt{\xi}}
     \nonumber \\
     \nonumber \\
     \K{8} & = & \kint y \nonumber \\
     & = &
     \K{1}+\left(1-\Frac{1}{2}\xi\right)\left[\,-\ln\xi+2\ln(2-\sqrt{\xi})-
     \ln\left({1+v\over1-v}\right)\,\right]
     \nonumber \\ &&
     -2(1-\sqrt{\xi})+v
     \nonumber \\
     \nonumber \\
     \K{9} & = & \kint {y\over z} \nonumber \\
     & = &
     \K{2}-\ln\xi+2\ln(2-\sqrt{\xi})-2\ln\left({1+v\over1-v}\right)
     \nonumber \\
     \nonumber \\
     v^3\,\K{10} & = & \kint {y\over z^2} \nonumber \\
     & = &
     2\left[\,\Li\left({1-v\over2-\sxi}\right)+\Li\left({2-\sxi\over1+v}\right)-
     2\Li\left(\sqrt{1-v\over1+v}\;\right)\,\right]
     \nonumber \\ &&
     +\ln\left({1+v\over1-v}\right)\left[\ln\left({1+v\over1-v}\right)-
     2\ln(2-\sxi)+\ln\xi-4v\right]
     \nonumber \\ &&
     +\ln(2-\sxi)\Big[\ln(2-\sxi)-\ln\xi+6v\Big]
     \nonumber \\ &&
     +\ln\xi\left(\Frac{1}{4}\ln\xi-3v\right)+4\left({1\over\xi}-{v^3\over\sxi}-1\right)
     \nonumber \\
     \nonumber \\
     \K{11} & = & \kint {yz} \nonumber \\
     & = &
     \K{8}-\Frac{1}{4}\left[10-9v+2v^2-v^3-2\sqrt{\xi}(4+\xi)\right]
     \nonumber \\ &&
     -\left(1+\Frac{1}{8}\xi^2\right)\left[\,\ln\left({1+v\over1-v}\right)
     +\ln\xi-2\ln(2-\sqrt{\xi})\,\right]
     \nonumber \\
     \nonumber
     \end{eqnarray}

The dilogarithm is defined by
$$  \Li(x) = -\int\limits_0^x{\ln(1-t)\over t},$$
and most of its known properties can be found in the standard reference 
Ref.~\cite{Lewin}.

Note that the soft divergences are at most logarithmic in $\Lambda$. In
$\K{7}$ all singularities cancel internally, and the finite part of
$\K{3}$ is denoted by $\KK{3}$. The integral $\S{12}$ has been calculated
in Ref.~\cite{kpt}.
\newpage
\newpage
\thispagestyle{empty}
\centerline{\bf\Large Figure Captions}
\vskip1cm
\newcounter{fig}
\begin{list}{
   \bf Fig.~\arabic{fig}:\ }{
         \usecounter{fig}
         \labelwidth1.6cm
         \leftmargin2cm
         \labelsep0.4cm
         \itemsep0ex plus0.2ex
        }
\item {\bf(a)} The longitudinal spin-spin correlation 
      $\langle P_{LL}\rangle$ at Born level and \oas\ for top pair 
      production in $e^+e^-\to\gamma,Z\to t\bar{t}$ as a function of
      $E_{cms}=\sqrt{q^2}$.
      {\bf(b)} Total \oas\ cross section for top-quark production as
      a function of the cms energy $E_{cms}=\sqrt{q^2}$.
      {\bf(c)} The impact of \oas\ corrections for the longitudinal top
      asymmetry $\langle P_{LL} \rangle$. Shown are values for
      $\langle P_{LL}(O(\alpha_s))\rangle/\langle P_{LL}(Born\/)\rangle-1$
      in percent.

\item {\bf(a)} QCD one-loop corrections of the longitudinal spin 
      correlations $\langle P_{LL} \rangle$ for 
      $\sigma_{tot}(e^+e^-\to\gamma,Z\to b\,\bar{b})$.
      {\bf(b)} \oas\ corrections for bottom pair production as a
      function of the cms energy $E_{cms}=\sqrt{q^2}$.
      {\bf(c)} \oas\ corrections compared to the Born contribution for
      the longitudinal spin correlation
      $\langle P_{LL} \rangle(e^+e^-\to\gamma,Z\to b\,\bar{b})$
      in percent.

\end{list}
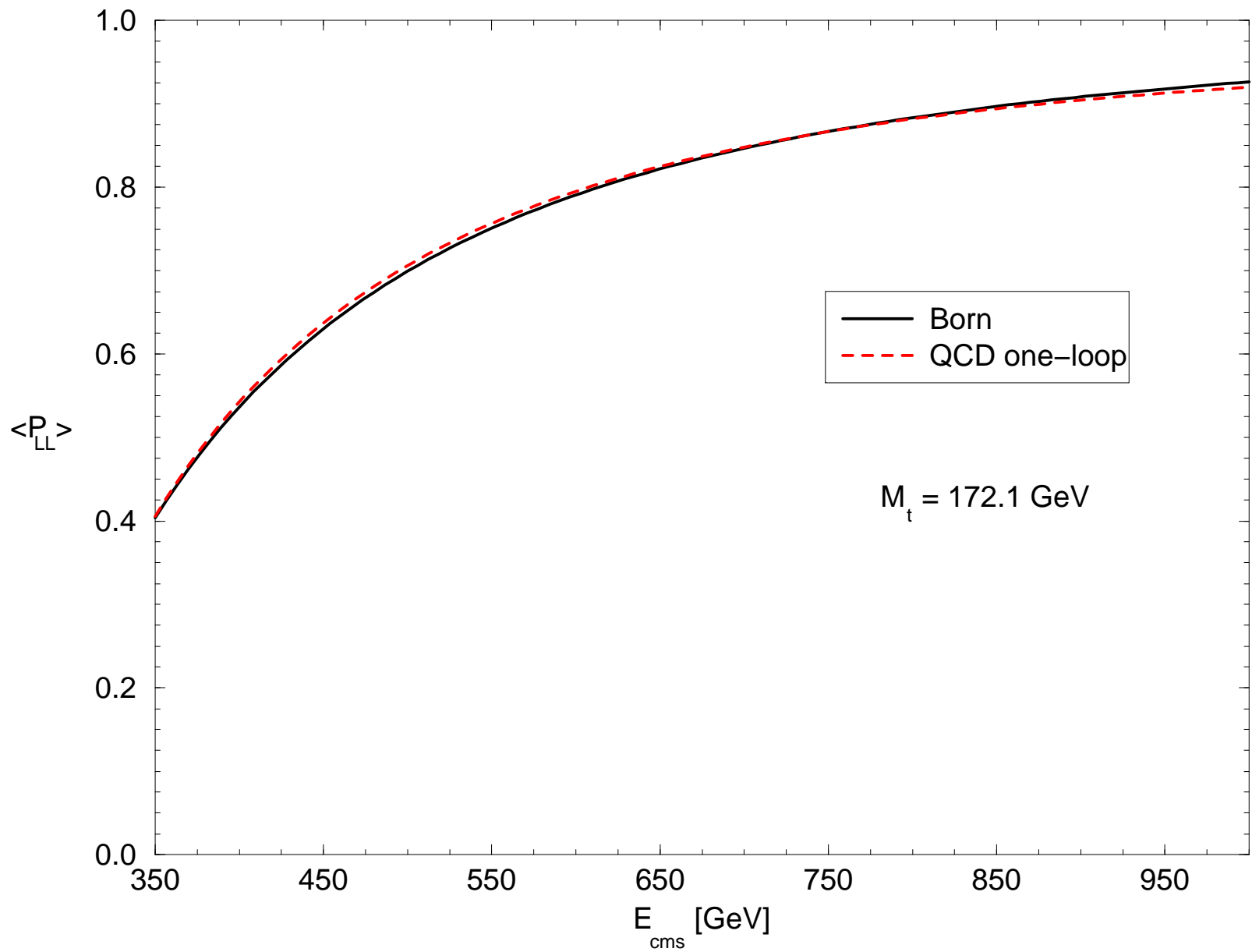
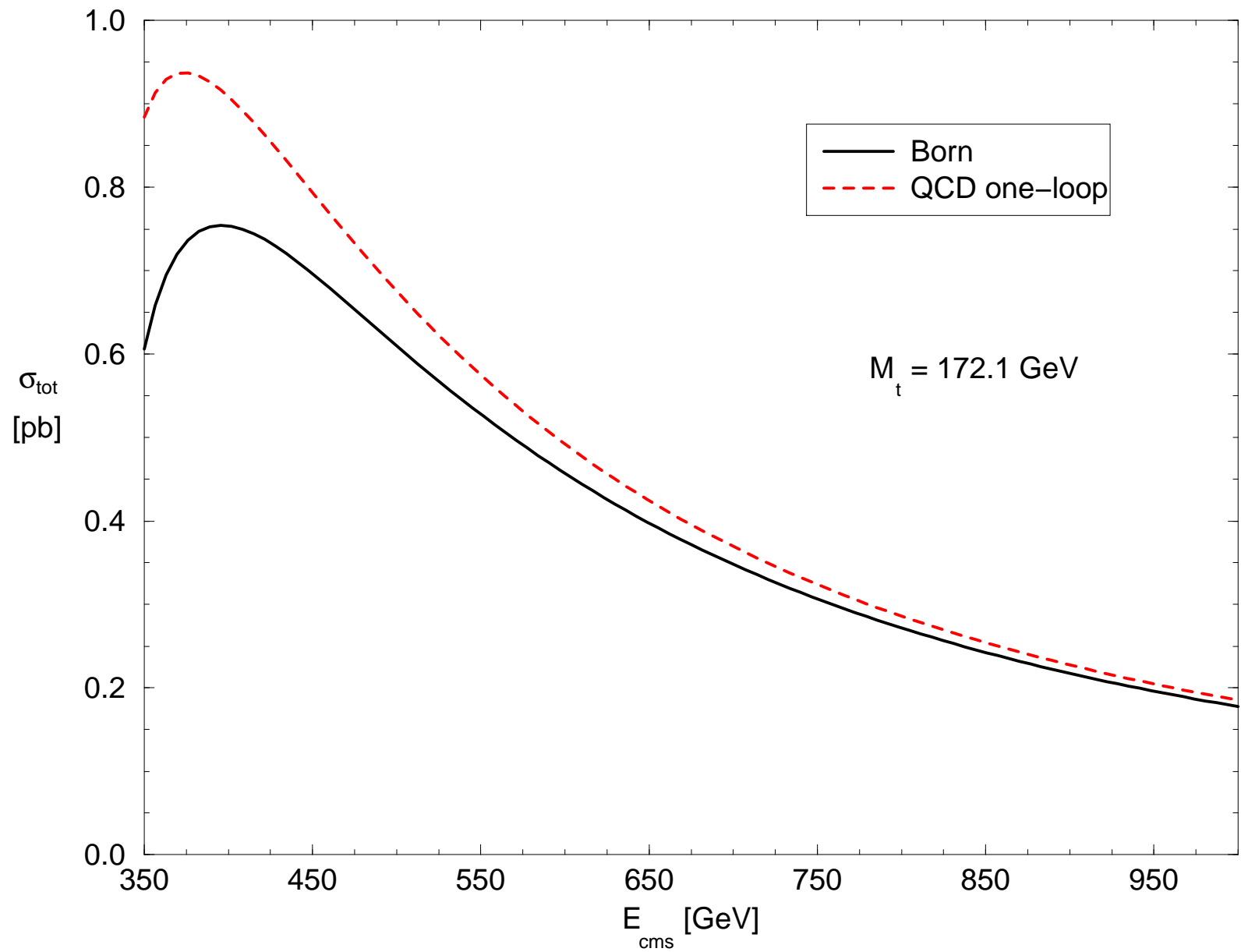
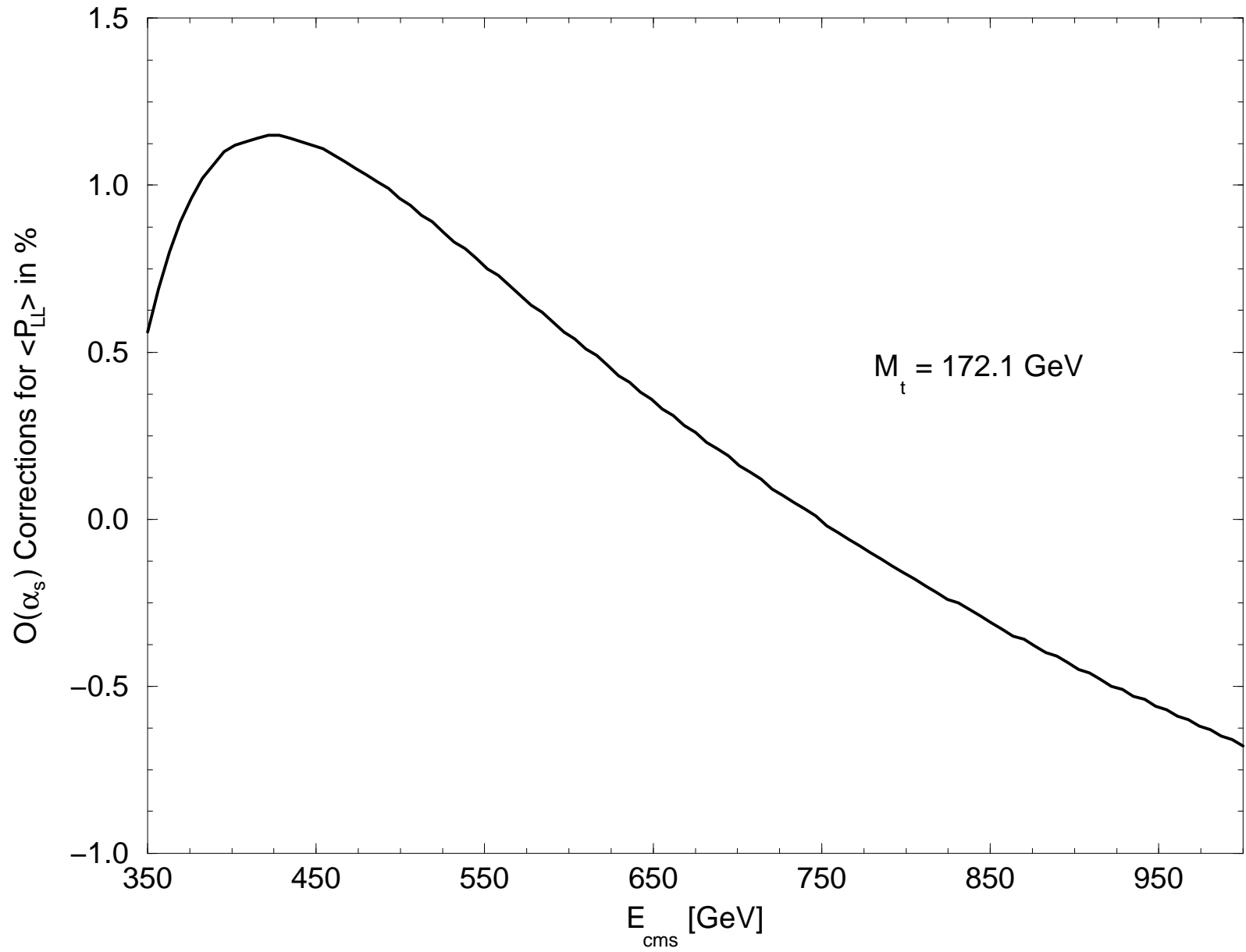
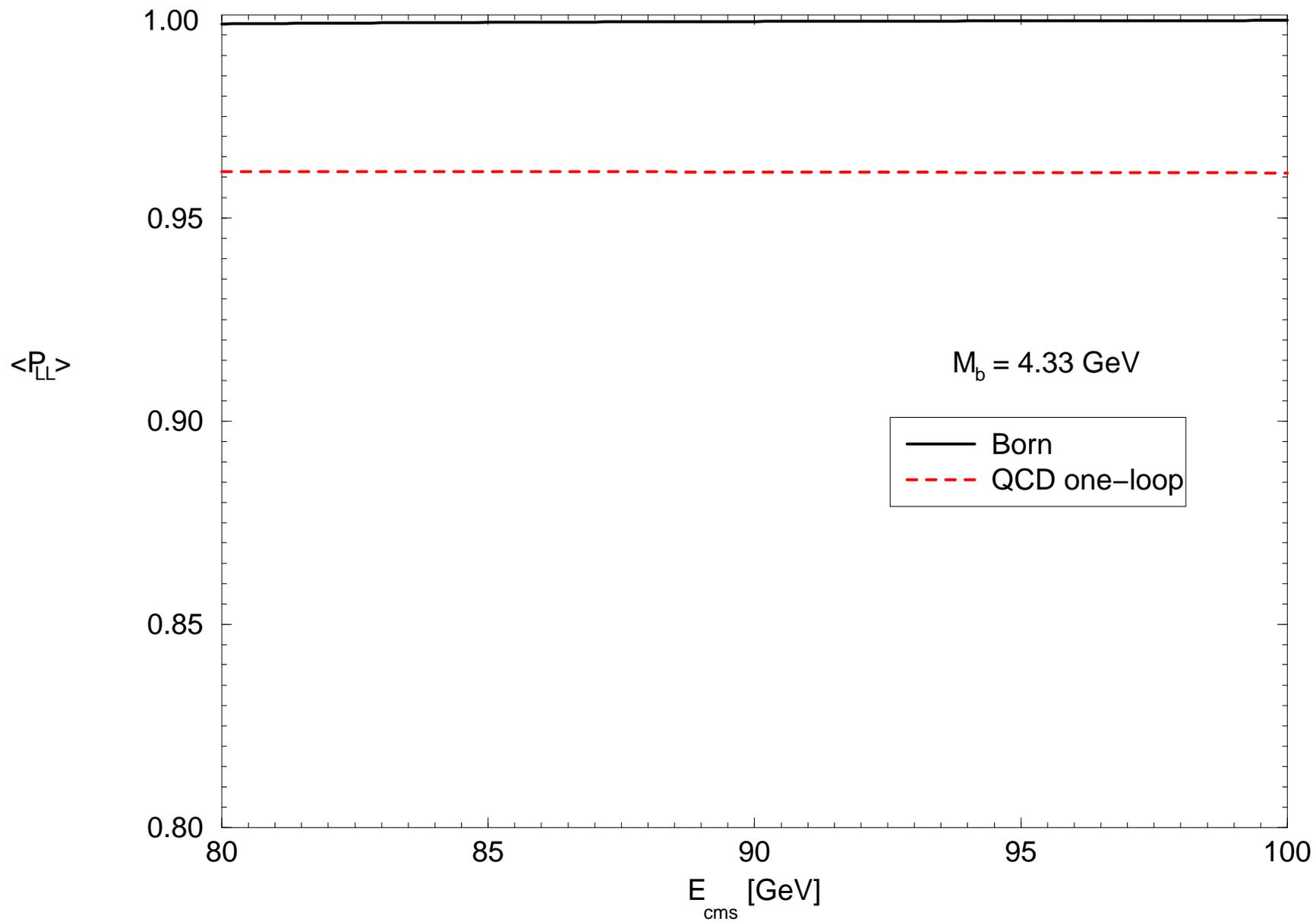
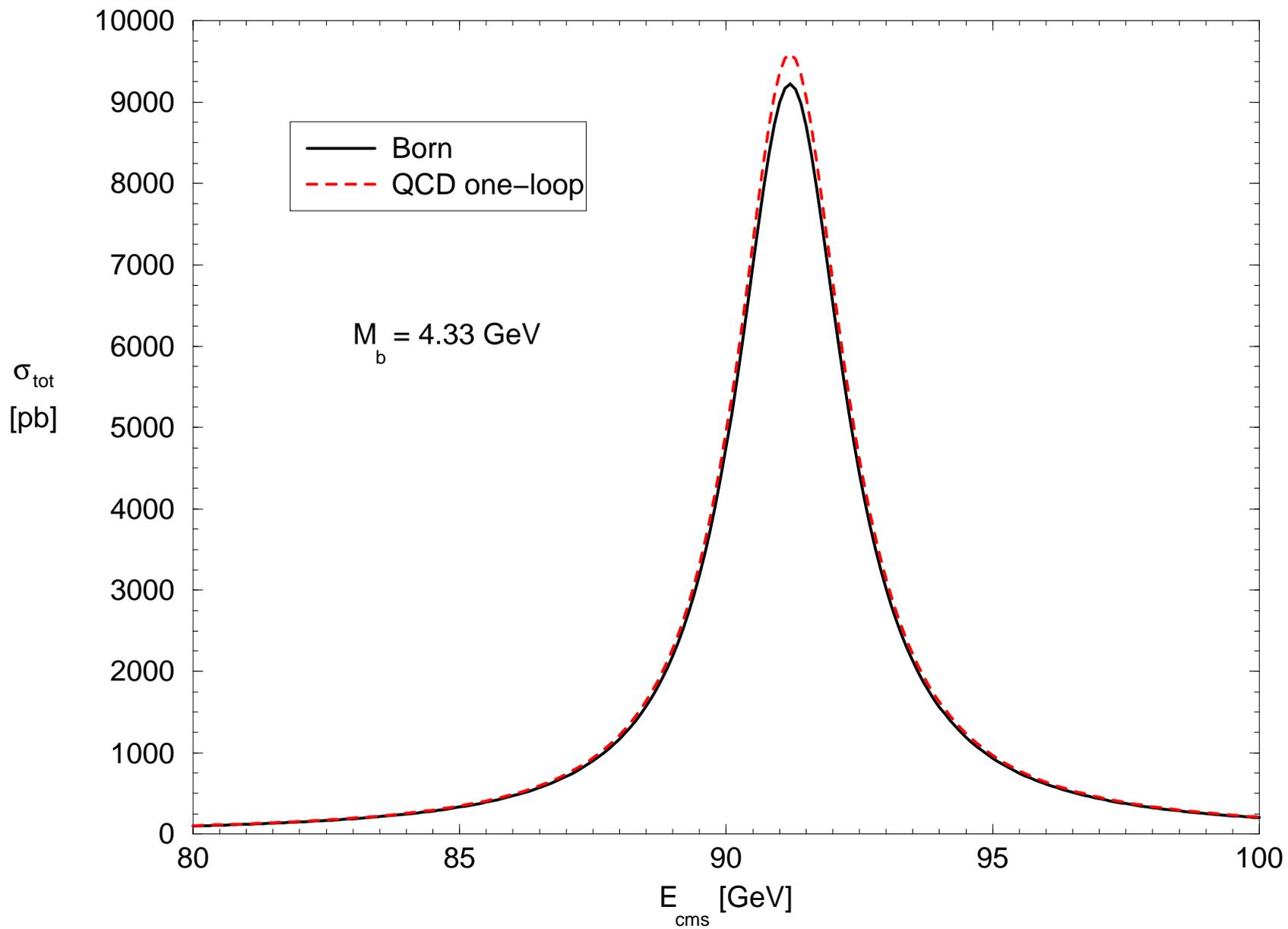
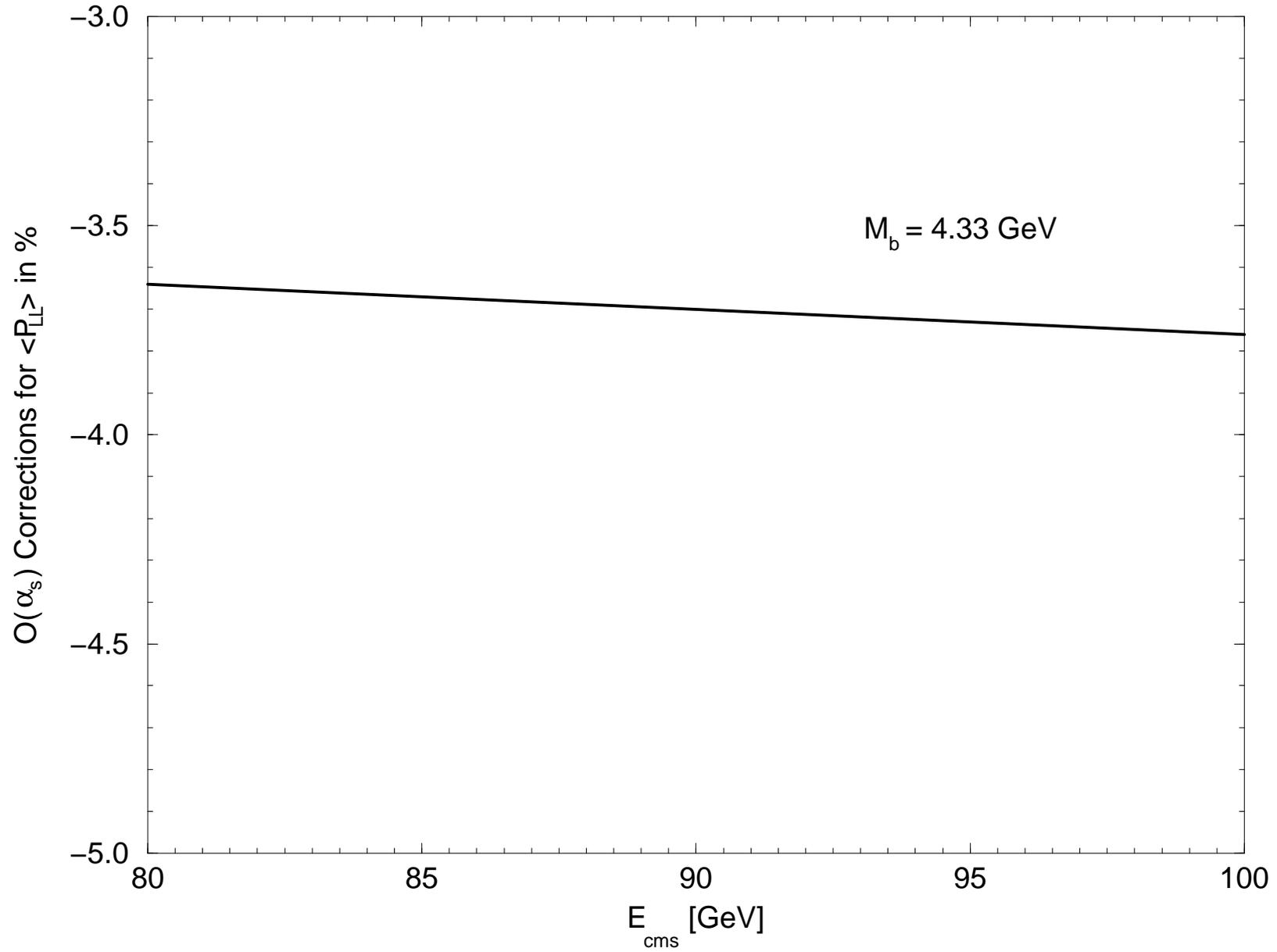
\end{document}